\documentclass[acmsmall, nonacm]{acmart}
\usepackage{longtable} 
\usepackage{graphicx} 

\usepackage{lineno}

\AtBeginDocument{%
  }

\copyrightyear{2025} 
\acmYear{2025} 
\acmDOI{XXXXXXX.XXXXXXX}

\acmJournal{JACM} 
\acmVolume{XX} 
\acmNumber{XX} 
\acmArticle{XX} 
\acmMonth{XX} 

\begin{document} 

\title[Sight, Sound and Smell in Immersive Experiences of Urban History]{Sight, Sound and Smell in Immersive Experiences of Urban History: Virtual Vauxhall Gardens Case Study}

\author[Tim Pearce]{Tim Pearce$^1$}
\affiliation{%
  \institution{School of Engineering, University of Leicester}
  \city{Leicester} 
  \country{UK} 
}
\authornote{Co-first authors}

\author[David Souto]{David Souto$^2$}
\affiliation{%
  \institution{School of Psychology and Vision Sciences, University of Leicester}
  \city{Leicester} 
  \country{UK} 
}
\authornotemark[1]

\author{Douglas Barrett$^2$}
\affiliation{%
  \institution{School of Psychology and Vision Sciences, University of Leicester}
  \city{Leicester} 
  \country{UK} 
}

\author{Benjamin Lok$^1$}
\affiliation{%
  \institution{School of Engineering, University of Leicester}
  \city{Leicester} 
  \country{UK} 
}

\author{Mateusz Bocian$^1$}
\affiliation{%
  \institution{School of Engineering, University of Leicester}
  \city{Leicester} 
  \country{UK} 
}
\authornote{Current address: School of Civil Engineering, University of Leeds, Leeds, UK}

\author{Artur Soczawa-Stronczyk$^1$}
\affiliation{%
  \institution{School of Engineering, University of Leicester}
  \city{Leicester} 
  \country{UK} 
}
\authornote{Current address: Bridge Engineering and Civil Structures Team, Buro Happold, London, UK}

\author[Giasemi Vavoula]{Giasemi Vavoula$^{3}$}
\affiliation{%
  \institution{School of Museum Studies, University of Leicester}
  \city{Leicester} 
  \country{UK} 
}

\author{Paul Long$^4$}
\affiliation{%
 \institution{MBD Limited, Leicester, UK}
 \city{Leicester} 
 \country{UK} 
}

\author{Avinash Bhangaonkar$^1$}
\affiliation{%
  \institution{School of Engineering, University of Leicester}
  \city{Leicester} 
  \country{UK} 
}

\author{Stephanie Bowry$^5$}
\affiliation{%
  \institution{School of Historical Studies, University of Leicester}
  \city{Leicester} 
  \country{UK} 
}

\author{Michaela Butter$^6$}
\affiliation{%
  \institution{Attenborough Arts Centre, University of Leicester}
  \city{Leicester} 
  \country{UK} 
}

\author{David Coke$^7$}
\affiliation{%
  \institution{Independent researcher}
  \city{London} 
  \country{UK} 
}

\author{Kate Loveman$^8$}
\affiliation{%
 \institution{School of Arts, Media and Communication, University of Leicester}
 \city{Leicester} 
 \country{UK} 
}

\author{Rosemary Sweet$^{9}$}
\affiliation{%
  \institution{School of History, Politics and International Relations, University of Leicester}
  \city{Leicester} 
  \country{UK} 
}

\author{Lars Tharp$^{10}$}
\affiliation{%
  \institution{The Foundling Museum} 
  \city{London} 
  \country{UK} 
}

\author{Jeremy Webster$^6$}
\affiliation{%
  \institution{Attenborough Arts Centre, University of Leicester}
  \city{Leicester} 
  \country{UK} 
}

\author{Hongji Yang$^7$}
\affiliation{%
  \institution{School of Computing and Mathematical Sciences, University of Leicester}
  \city{Leicester} 
  \country{UK} 
}

\author{Robin Green$^2$}
\affiliation{%
  \institution{School of Psychology and Vision Sciences, University of Leicester}
  \city{Leicester} 
  \country{UK} 
}

\author{Andrew Hugill$^{7,11}$}
\affiliation{%
  \institution{Institute for Digital Culture \& School of Computing and Mathematical Sciences, University of Leicester}
  \city{Leicester} 
  \country{UK} 
}
\authornote{Corresponding author: ah619@leicester.ac.uk}

\begin{abstract}
We explore the integration of multisensory elements in virtual reality reconstructions of historical spaces through a case study of the Virtual Vauxhall Gardens project. While visual and auditory components have become standard in digital heritage experiences, the addition of olfactory stimuli remains underexplored, despite its powerful connection to memory and emotional engagement. This research investigates how multisensory experiences involving olfaction can be effectively integrated into VR reconstructions of historical spaces to enhance presence and engagement with cultural heritage. In the context of a VR reconstruction of London's eighteenth-century Vauxhall Pleasure Gardens, we developed a networked portable olfactory display capable of synchronizing specific scents with visual and auditory elements at pivotal moments in the virtual experience. Our evaluation methodology assesses both technical implementation and user experience, measuring presence, and usability metrics across diverse participant groups. Our results show that integrating synchronized olfactory stimuli into the VR experience can enhance user engagement and be perceived positively, contributing to a unique and immersive encounter with historical settings. While presence questionnaires indicated a strong sense of auditory presence and control, with other sensory factors rated moderately, user experience of attractiveness was exceptionally high; qualitative feedback suggested heightened sensory awareness and engagement influenced by the inclusion and anticipation of smell. Our results suggest that evaluating multisensory VR heritage experiences requires a nuanced approach, as standard usability metrics may be ill-suited and 'realism' might be less critical than creating an evocative, historically informed, and emotionally resonant experience. This research contributes to the growing field of multisensory museum technologies, offering practical insights for cultural institutions seeking to create more inclusive and immersive exhibitions. The findings suggest important implications for future heritage installations, particularly in terms of accessibility for diverse audiences, technical integration challenges, and the development of historically authentic sensory experiences that can preserve intangible aspects of cultural heritage that might otherwise be lost.
\end{abstract}

\renewcommand{\shortauthors}{Pearce et al.}




\maketitle

\section{INTRODUCTION AND PROJECT OVERVIEW}
\label{sec:introduction}
Virtual reality (VR) has revolutionized digital experiences, yet most applications primarily engage only the visual and auditory senses, leaving the olfactory system largely untapped, despite its profound connections to memory and emotion. Our research investigates how multisensory experiences, particularly olfaction, can be effectively integrated into VR reconstructions of historical spaces to enhance the presence and engagement with cultural heritage. This question addresses a significant gap in current VR research, as studies suggest that olfactory stimuli combined with visual environments can significantly increase presence and trigger stronger emotional responses and memory recall [1]. The integration of appropriate scents—whether the musty aroma of ancient manuscripts, the smoke-laden atmosphere of industrial workplaces, or the culinary scents of historical kitchens—presents unique technical and experiential challenges, but offers the recreation of more authentic, memorable, and emotionally resonant encounters with the past [2].

Although traditional approaches to history prioritize textual knowledge, advances in immersive VR technology have created new opportunities to recapture lost sensory experiences. Our perception of history has been shaped primarily through documents, artifacts, and visual representations, yet the full sensory reality of historical environments remains largely inaccessible to us. The Virtual Vauxhall Gardens (VVG) project focused on creating an accurate reconstruction of the Rotunda – the main indoor entertainment hub of the gardens – based on archival sources and pictorial representations. We placed particular emphasis on recreating the complete sensory environment visitors would have experienced, including the technical challenges of scent selection, odor delivery timing, and congruence with visual stimuli to explore future frameworks for cultural heritage experiences that engage multiple sensory pathways. We developed a networked portable olfactory display system capable of synchronizing specific scents with visual and auditory elements at pivotal moments, to demonstrate how sensory expansion beyond audiovisual stimuli can create more deeply felt and memorable encounters with otherwise inaccessible sensory dimensions of the past. The VVG recreation was first exhibited at Attenborough Arts Centre, Leicester on July 19-20, 2019 and subsequent user experiences were evaluated through focus groups and standardized questionnaires measuring presence, enjoyment, and usability.

This paper examines our multisensory approach through nine interconnected sections. Following this introduction, establishing our research context, we present the historical significance of Vauxhall Gardens as an ideal test case for multisensory reconstruction due to its deliberate design as a space of sensory pleasure and transformation. The technical methodology section details our approach to translating historical documentation into a functioning VR environment, comprising a bespoke olfactory display that enables precise scent delivery timed to specific virtual events. The findings of the focus group evaluation reveal how the participants responded physiologically and emotionally to the multisensory experience. The discussion examines how this case study contributes to broader questions about historical sensory recreation, addressing both the limitations of modern sensory expectations when encountering historical stimuli and the implications of the case study for future multisensory VR heritage installations. Throughout, we emphasize the interdisciplinary collaboration that enabled this research, suggesting a model for future work integrating sensory history with technological innovation.

\section{HISTORICAL CONTEXT OF VAUXHALL GARDENS}
\label{sec:historical_context}
The area that became Vauxhall Gardens had an inauspicious beginning. From the site of an out-of-town tavern frequented by young city gentlemen in their leisure hours, and consequently by prostitutes working in London's sex trade, New Spring Gardens on the south bank of the Thames in London's suburb of South Lambeth became one of the great public entertainment venues of the 20th-century. Opened in the euphoric times following the Restoration of 1660, it was, in 1732, transformed into ‘Vauxhall Gardens’, and re-branded as a fashionable commercial Pleasure Garden, by a remarkable young entrepreneur named Jonathan Tyers. On property which was within the London estates of the Prince of Wales, Tyers created a cultural site where Londoners and visitors could enjoy a polite evening out with music and refreshments available away from the cares of everyday life and work.

Vauxhall Gardens became a prominent pleasure garden in 18th-century London, playing a crucial role in the city's social and cultural life. Situated on the south bank of the Thames, it was a part of the urban environment that was marketed as an escape from the everyday sights and smells of the city.

In the mid-18th century, Tyers transformed Vauxhall into a magical, enclosed wooded garden, marketed as an escape from the city's everyday sights and smells, where visitors felt transported to an exotic, other-worldly space. This re-creation offers opportunities to enhance understanding of the capital's historical environment and fosters creative engagement with the urban past. Every aspect of Tyers's new pleasure garden was specifically designed to reinforce this impression, from the anticipatory journey across the river by rowing boat, to the sights and sounds, tastes and smells to be experienced on arrival. The overriding pleasure that visitors gained from Vauxhall was sensory – in stark contrast to the dangerous, malodorous and cacophonous streets of London where all senses had to be suppressed, Vauxhall gave its visitors the freedom to open up their senses, to accept the refreshing stimuli, and to let down their habitual guard and actually relax and enjoy themselves [3]. With its revolutionary open-air bandstand, its 'Umbrella Room' or Rotunda, and its Chinese and Turkish pavilions, Vauxhall Gardens was open on summer-time evenings after work and became the most popular and influential of all the pleasure gardens of Georgian London, visited by everybody from the royal family to apprentices and servants.

Advertising and contemporary accounts confirm Tyers offered an all-embracing sensual experience that captivated visitors. They reportedly departed elated, senses tingling and feeling revitalized, a pleasurable stimulation that encouraged many return visits [4].

The enchanted atmosphere of the gardens was heightened by the architecture, the promenades, the paintings and sculpture installed there by Tyers, with the advice of his friend and artist William Hogarth. Tyers had been in trade in Bermondsey before taking the lease of Vauxhall, and he was a skilled and astute businessman. He also knew how to delegate in areas outside his expertise; the influence of Hogarth was obvious in the strong visual arts presence at Vauxhall, but Tyers was also able to involve the outstanding musician and composer G.F. Handel, whose influence on the music performed nightly at the gardens was profound and long-lasting. Both Hogarth and Handel would have known the importance of the high quality that became Tyers's watchword in everything he did. The final element in the creation of the classic English pleasure garden was provided by an unlikely participant; John Lockman, poet and translator, who brought to Vauxhall a marketing genius and that influenced all aspects of the gardens, from the relatively cheap one shilling admission price to the installation of exotic architecture, the inclusion of popular song in the music programs, and protection from the always unreliable English weather [5].

\begin{figure} 
  \centering
  \framebox[0.8\textwidth][c]{\includegraphics[width=0.75\textwidth]{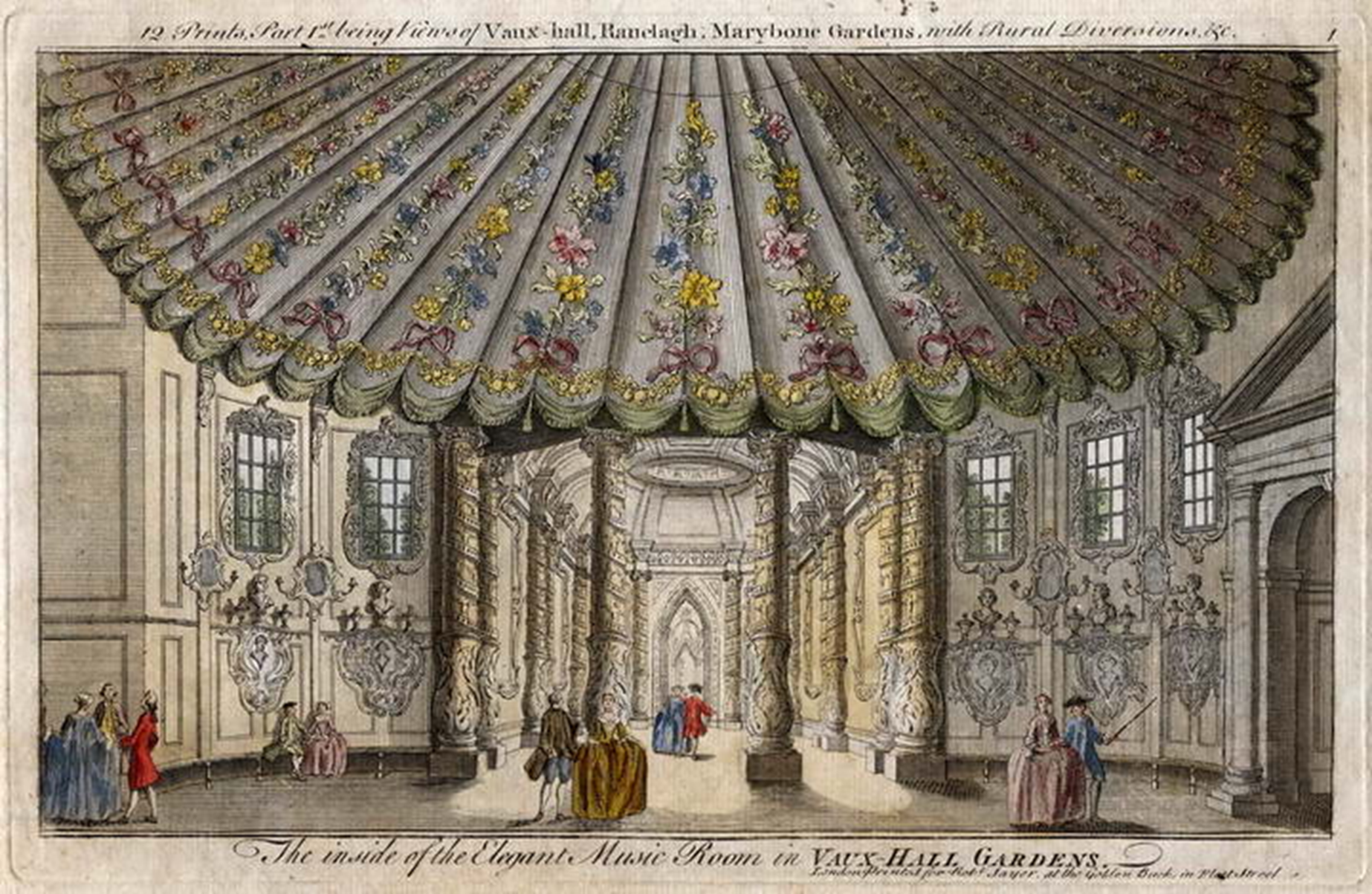}} 
  \caption{The Inside of the Elegant Music Room in Vaux-Hall Gardens' [the Rotunda] (anonymous engraving after B. Cole, c.1752. David Coke's collection CVRC 0127, with permission).}
  \Description{An 18th-century engraving showing the ornate interior of the Vauxhall Gardens Rotunda, filled with elegantly dressed people, musicians in a gallery, and elaborate decorations.}
  \label{fig:vauxhall_music_room}
\end{figure}

After being run as a hugely successful business by the Tyers family for a century and subsequently passing through the hands of several different owners, Vauxhall Gardens finally closed in the summer of 1859; the valuable site was soon cleared and developed for housing. Many artefacts from the gardens survive today, and many pieces of visual documentation too. However, although we can reconstruct the look of the gardens with relative confidence, it is remarkably difficult in the twenty-first century to reconstruct the wonder and astonishment felt by first-time visitors, as well as the almost mind-altering sensory response that people likely experienced at the time, which not only enriched their visit, but encouraged them to return to Vauxhall again and again, and to tell their friends about it [6].

\section{METHODOLOGY: VR RECONSTRUCTION}
\label{sec:methodology_vr_reconstruction}
The architectural model of the Rotunda, Vauxhall Gardens' main entertainment hub, was developed in ArchiCAD, optimized for Unreal Engine, and textured to simulate stone, plaster, gold gilt, and painted surfaces. Details like busts, urns, carvings, and the 72-candle chandelier were added, and the model was lit with realistic shadows and surface textures, including flickering candles, to enhance immersion and bring the building to life. The VR experience was delivered via Oculus Rift S headsets. Readers can get an impression of the virtual experience through the project website.\footnote{Project website \url{https://andrewhugill.com/virtualvauxhallgardens/}, accessed 28 February 2025), including a video walkthrough of the experience (\url{https://youtu.be/-qSfxeVZP_g8}).}

The sources on Vauxhall Gardens in the 1740s and 1750s range from newspaper articles, advertisements, and song-sheets, to diaries, poems, and novels.\footnote{A full list of these sources is given in D. Coke and A. Borg, Vauxhall Gardens: A History (Yale University Press, 2011), 453; and on Coke's website \url{http://www.vauxhallgardens.com/} (accessed 29 April 2020) under 'Contemporary Sources'.} For the VR model, the most important survivals were the visual sources – engravings of the Rotunda – and those narrative accounts which convey the experience of visiting the gardens. Each of these sources posed problems for reconstruction. The visual references for the Rotunda in the mid-century are all based on one drawing by Samuel Wale that was reproduced in 1752 as a print by H. Roberts titled ‘The Inside of the Elegant Music Room in Vaux Hall Gardens' (Figure~\ref{fig:vauxhall_music_room}).

This print survives in huge numbers, with a bewildering variety of hand-coloring, almost all of which is unreliable. To determine the color scheme of the Rotunda, scattered anecdotal references in the sources therefore had to be combined with educated conjecture: the final scheme was based on probability (ceilings were often painted blue) and on the known occurrence of colors in surviving similar spaces (in which, for example, window-frames were often gilded). Meanwhile, narrative sources—guidebooks, magazine articles or poems which portray visits to Vauxhall Gardens—tended either to be designed to promote the attraction to visitors or to satirize those same visitors. The accuracy of these accounts, therefore, had to be carefully gauged. These texts did, nonetheless, indicate shared itineraries, on which we drew to create a structure for the audience's progress through the virtual reconstruction.

\begin{figure*} 
  \centering
  \includegraphics[width=0.5\textwidth]{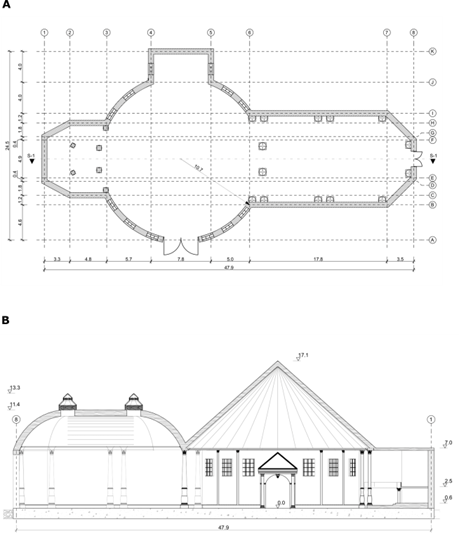} \\ 
  \vspace{0.5cm}
  C \\
  \includegraphics[width=0.6\textwidth]{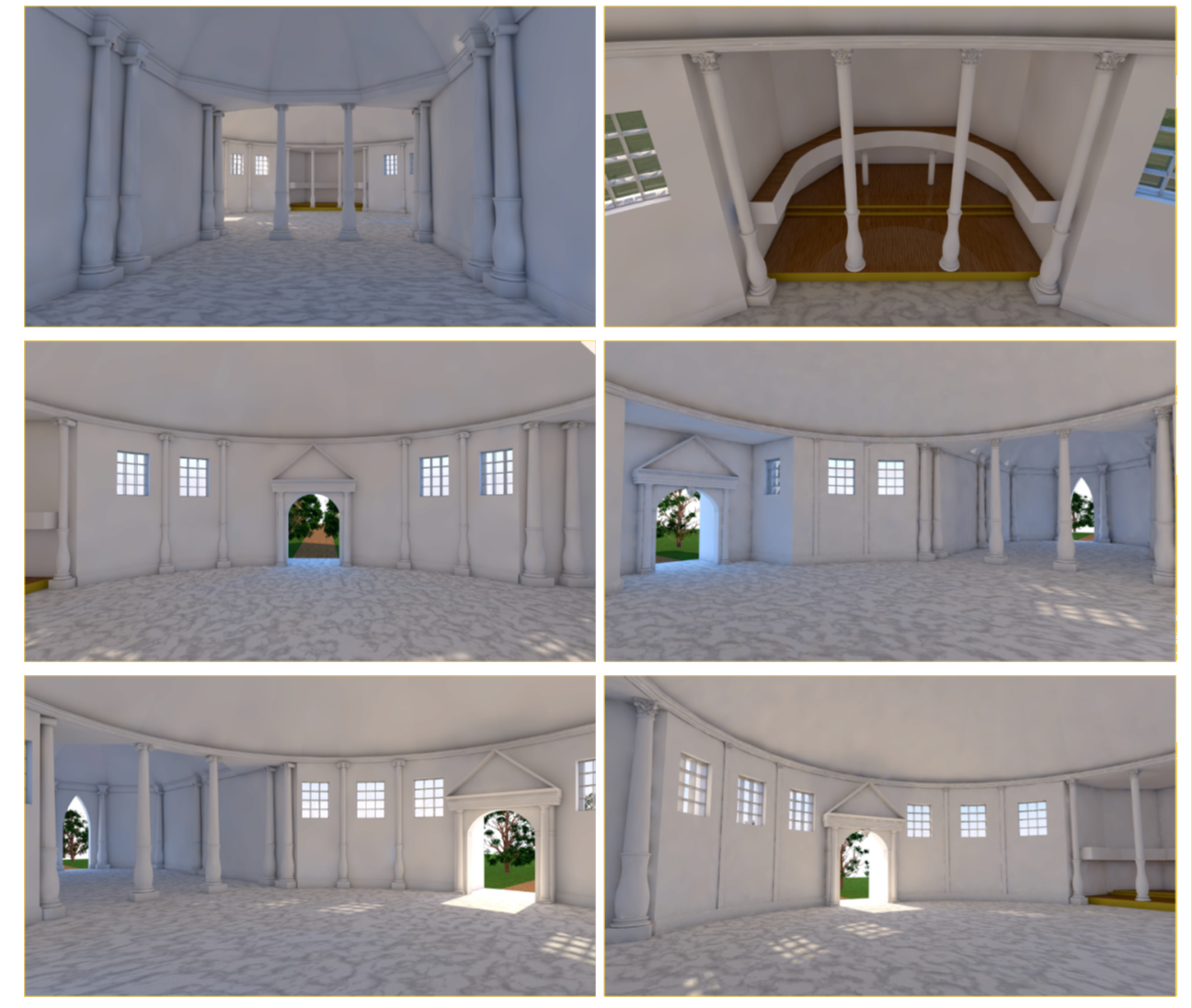} \\ 
  \caption{Aerial plan of the Rotunda (A) and side section (B). Another example of those discrepancies was the positioning of columns supporting the main rotunda's roof next to the stage area. The plan drawing suggests that the columns were distributed along a straight line, which was deemed unrealistic due to the fact that their main function was to transfer the weight of the rotunda down to the foundations. To address this issue, the columns in the virtual model were distributed along a circular path mapping the outline of the rotunda, thus ensuring structural integrity (C) Interior architectural virtual model before decoration (artwork: Paul Long, Artur Adam Soczawa-Stronczyk \& Supervision: Mateusz Bocian).}
  \Description{Figure 2 shows: (A) An architectural top-down plan of the Rotunda. (B) A side elevation architectural drawing of the Rotunda. (C) Six rendered images of the interior architectural virtual model of the Rotunda before decoration, showing white walls, columns, and arched doorways/windows.}
  \label{fig:rotunda_plans_sections}
\end{figure*}

An initial architectural model was prepared in ArchiCAD building information modelling (BIM) software. During the modelling process, some initial assumptions had to be made on the geometry and the proportions of the structure, as many sources were found to provide contradictory information or only offered artistic representation, disregarding the physical constraints of the structure itself. The sizes of the main rotunda section and the Pillared Saloon attached to it had to be reconciled. The contemporary descriptions indicate that those sections were of similar area, as it was said that the Pillared Saloon doubled the audience capacity of the Rotunda. However, according to the drawing, the main rotunda section had an area of approximately 318 m$^2$, whereas the Pillared Saloon was smaller by nearly 130 m$^2$. Consequently, a decision was made to extend the Pillared Saloon in the model by over 4 meters to increase its area while keeping the proportions of the whole building as consistent with the plan drawing as possible (Figure~\ref{fig:rotunda_plans_sections}).

Once the model was completed, we generated the narrative structure to create vignettes that covered key themes from the research. Sections covered the construction of the building itself, the music that was presented in the rotunda, and the thinking behind Jonathan Tyers' ambitious vision. Written text was compiled, edited and recorded before being set to an atmospheric musical backdrop.

\section{EXHIBITION DESIGN AND USER EXPERIENCE}
\label{sec:exhibition_design}
Since the installation was targeted at museums, various limitations were decided upon from the start. Immersion was visually limited, with the head-mounted display covering only 90 degrees (horizontally) of the 200 degrees of our normal binocular visual field [7]. Visitors were moved through the space without user input, using fades and transitions, rather than encouraging teleportation. The addition of 'view to activate' icons meant that viewers could (and usually did) complete the whole experience without being given controllers at all, or any ability to teleport/move through space. This made it significantly easier to use for people who have little to no experience with VR, and easier to explain for staff or stewards helping people through.

The exhibition had three components: an interactive VR experience delivered via a headset while seated; a large projected slideshow providing historical background about the Gardens; and a series of A0 posters describing the contributions of each of the disciplinary partners and the project as a whole. There were three VR headsets, one of which was equipped with the olfactory module described below. The other components of the exhibition occupied and informed people while they waited for a headset to become available. The exhibition took place in the University of Leicester Attenborough Arts Centre – across two days, hosting approximately 100 visitors. The visitors included a wide age range, from children to elderly people, and a similarly diverse range of backgrounds and interests. Several had travelled long distances to attend and there were representatives from the Vauxhall Trust and the Friends of Vauxhall Gardens, two groups active in promoting the culture and history of the current site. Perhaps the most extraordinary visitor was a direct descendant of Jonathan Tyers himself.

For visitors to the exhibition, the VR experience began with the river-crossing. This was a vital part of a visit to Vauxhall in the mid 18th-century, when London Bridge was the only terrestrial crossing available. In the recreation, pile moorings and impressionistic images of the water are accompanied by an introductory voice-over. The user hears a female voice reading adapted excerpts from an eighteenth-century source, followed by a male voice who offers further detail on what the user is hearing and seeing: this pattern of narration is followed in all sections of the experience.\footnote{The voiceovers are by Paige Emerick and Lars Tharp.} Our principal source for the female voiceover was an article in The Scots Magazine in 1739 [8]. This described for its Scottish readership how attending Vauxhall worked, starting with a journey there made by a mixed group of excited visitors. The first port of call at the gardens, if the weather was inclement, would have been the Rotunda, just on the left after the entrance. Therefore, as our user arrives, the voiceover announces rain, and the group rushes to the Rotunda. The Rotunda then rises around the user. During daylight hours, the gardens and trees outside are visible through the doors and windows. Darkness soon falls, accompanied by one of the great and memorable special effects of Vauxhall – the illuminations.

The giant chandelier that lit the Rotunda was one of the known lacunae in the extant sources. Due to its overpowering size, this chandelier was never included in the topographical prints of the interior of the room, so its exact form is unknown. Descriptions of the chandelier tell us that there were three rows of candles, 72 in all, and that it was eleven feet in diameter, with the arms of the chandelier arranged around central spheres [9]. This led to the conclusion that it resembled the Flemish-style brass examples often used to illuminate churches, so these were adopted as our pattern. A famous special effect at Vauxhall was the lighting ‘in an instant' of all the lamps (inverted glass ‘bells' with oil-lamps inside, fueled by whale-oil), which was achieved using long flammable fuses running from lamp to lamp [10]. It was deemed important that the virtual reconstruction should represent this spectacular effect as faithfully as possible. Users, therefore, could interact with the model to light the chandelier and even smell the burning lamps, and, in an additional section, hear the music played by Vauxhall's band, which would have entertained visitors indoors on rainy days.\footnote{The music used was Handel's “The Advice: Mortals wisely learn to measure”. Sophie Bevan (soprano), London Early Opera, Bridget Cunningham (conductor). Available on the album 'Handel at Vauxhall 1' SIGCD428. Used with permission.}

After this, and accompanied by the final voice-over from The Scots Magazine, the scene darkens, and the viewer departs. The whole ‘visit' has only lasted a few minutes, but it aimed at placing the user actually within the three-dimensional space of a building that was demolished in 1859, capturing for modern visitors something of what it might have felt like to be there in the 1750s. It also aids historians in visualizing this remarkable space in real time and in understanding the powerful combination of sensory impressions that must have made it so extraordinary an attraction.

\section{OLFACTORY EXPERIENCE: THEORY AND IMPLEMENTATION}
\label{sec:olfactory_experience}
One distinctive feature of the exhibition was the inclusion of smell. Time has, of course, shifted our olfactory sensibilities. In the eighteenth century, particular odors would have been pervasive, rather than objects of distant ‘otherness'. Indeed, a visitor to mid-eighteenth-century Vauxhall, quitting the everyday stench of London, would probably have been struck by the lack of stink. Far more arresting would have been the art and the sounds. But for present-day audiences in a VR environment, the inclusion of the smell of the burning candles in the chandelier could have contributed to a sense of otherworldly excitement. The recreation of a 'pleasurable' and 'sensational' historical experience is not literal, but must adapt to the requirements of its time.

Including the olfactory modality was challenging for several reasons: the ephemeral nature of odor renders physical evidence from the past largely unavailable, so that direct evidence is generally lost to us; although typically considered a “slow” sense, olfactory stimuli are invisible and fleeting; modern olfactory language is strangely limited, generally relying on culturally relative metaphors that limits our ability to specify and compare odors—in other words smell is linguistically challenged [11]. Odor identity and meaning are also highly contextual, strongly determined by specific cultures, times and places.

This historical demotion of olfaction, alongside the challenges detailed above, likely explains its systematic neglect in many historical exhibits.

\begin{figure} 
  \centering
  \includegraphics[width=0.7\textwidth]{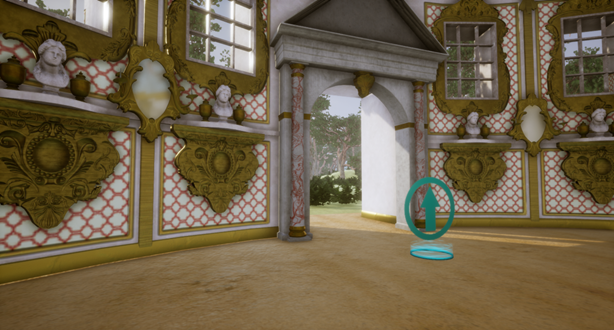} \\ 
  \vspace{0.5cm}
  \includegraphics[width=0.7\textwidth]{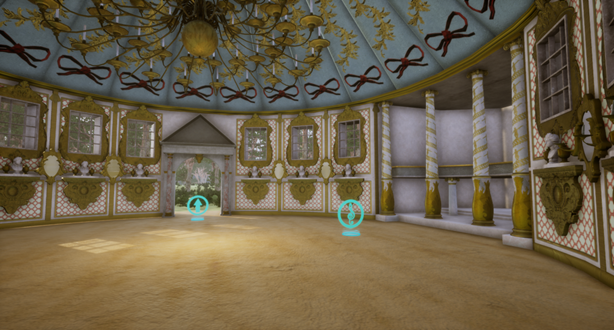} \\ 
  \caption{Stills from VR Vauxhall Gardens within the Rotunda including candelabra (bottom).}
  \Description{Two stills from the VR experience. Top  view towards the entrance of the Rotunda. Bottom: centre of the Rotunda with candelabra.}
  \label{fig:vr_stills}
\end{figure}

\subsection{VR Olfactory Display Technical Implementation}
\label{ssec:olfactory_technical}
\begin{figure} 
  \centering
  \includegraphics[width=1\textwidth]{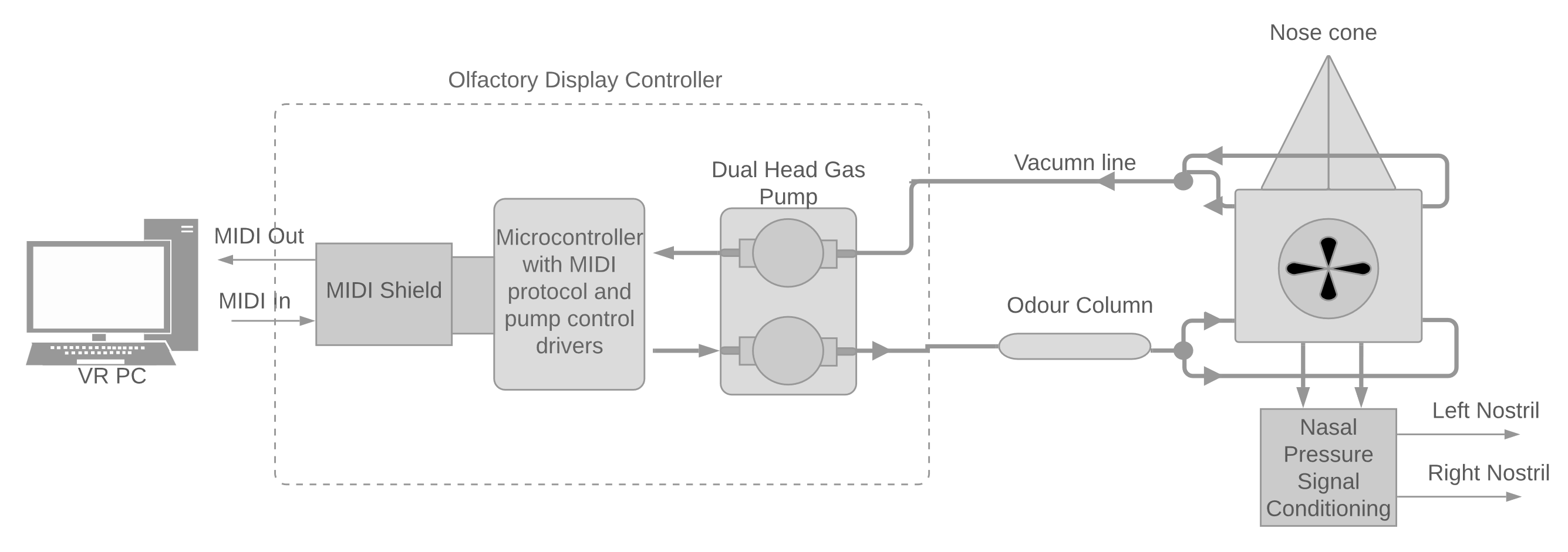} 
  \caption{University of Leicester Olfactory Display (ULOD) system schematic.}
  \Description{A schematic diagram showing components of the olfactory display system: VR PC connected to an Olfactory Display Controller (Microcontroller, Dual Head Gas Pump), which leads to an Odor Column and then to a Nose cone with connections for Left Nostril, Right Nostril, and Nasal Pressure Signal Conditioning.}
  \label{fig:ulod_schematic}
\end{figure}

For this exhibition, we had to overcome various challenges to reliably and safely deliver olfactory stimuli in the context of a public VR installation. A schematic of the University of Leicester Olfactory Display (ULOD) is shown in Figure~\ref{fig:ulod_schematic}.

\begin{figure*} 
  \centering
  \includegraphics[width=0.9\textwidth]{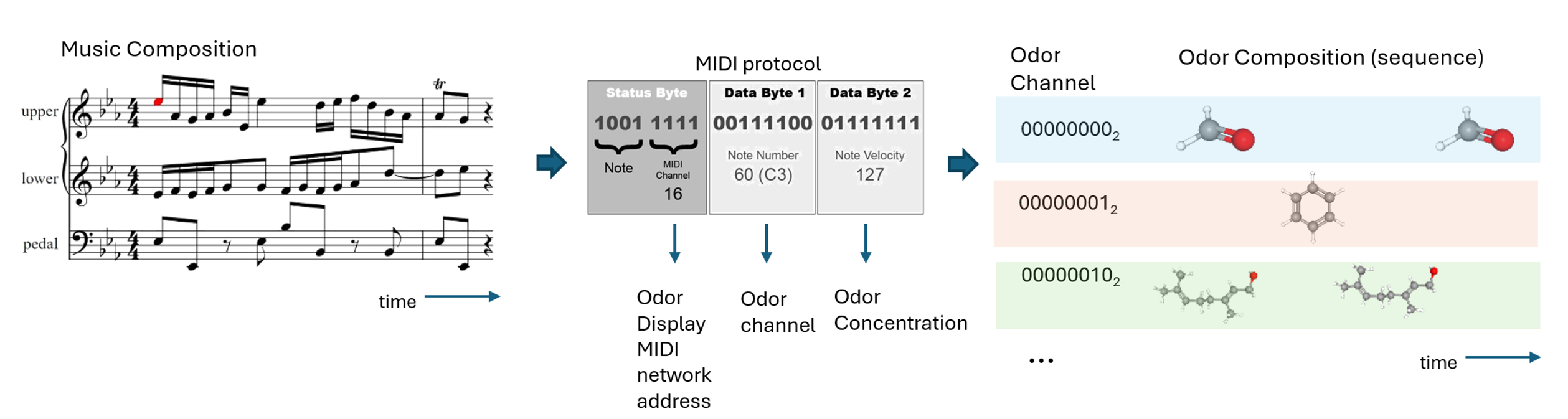} 
  \caption{Our olfactory display repurposes the MIDI protocol to support multi-channel odor compositions in time, synchronized with VR events. Olfactory displays can be daisy-chained to add additional odor channels.}
  \Description{A diagram illustrating how a music composition is translated into MIDI protocol, which then controls an odor channel to produce an odor composition sequence over time. Arrows show data flow from music notation to MIDI bytes (Note On, MIDI Channel, Note Number, Note Velocity) and then to Odor Display (MIDI network address, Odor channel, Concentration), finally resulting in a timed sequence of odor release represented by chemical structures.}
  \label{fig:midi_olfactory}
\end{figure*}

There are multiple technical challenges related to odor delivery. Specifically, in its development, we needed to consider:
\begin{enumerate}
    \item[a)] Avoiding absorption onto the equipment materials with the volatile compounds to reduce contamination effects. We addressed this by using PTFE/silicone materials where possible in the design, in particular the odor column and tubing to the mask, to make it resistant to odor absorption.
    \item[b)] User safety in terms of limiting chemical and electrical exposure, working with a range of sensory thresholds that vary across individuals substantially (up to 5-6 orders of magnitude) and stability of concentration over time. The design ensures that no electrical components came into contact with the user at any time. Using a dual-head digital pump with speed control ensured that the chemical stimulus could be calibrated to a specific concentration.
    \item[c)] Ability to accurately initiate and terminate the olfactory signal and achieve synchronisation with VR content. To achieve this we repurposed the MIDI protocol used for music data transmission as a means to control the olfactory display from the VR engine (Figure~\ref{fig:midi_olfactory}). This provides the possibility to programmably deliver sequences of olfactory stimuli as a composition to match a specific VR experience, making it flexible for wider use.
    \item[d)] Silent operation to avoid auditory cues. We used sound-damping materials inside the olfactory display to minimise pump noise and this was measured to be close to the threshold of human hearing at 1 metre distance.
    \item[e)] Portability and low power. The olfactory display includes a power management system to deliver power from a rechargeable battery for portability that can run for at least one day of demonstrations before recharge.
    \item[f)] User comfort that does not interfere with other sensory modalities, so that it remains largely invisible to the user. Our 3d-printed olfactory display mask was designed for human acceptability and direct coupling to the Oculus Rift S. The mask optionally includes left/right nasal pressure monitoring of the breathing/sniff cycle to assess user interaction with the olfactory display technology.
    \item[g)] Low-cost. For the Vauxhall Gardens installation, we limited the design to a single olfactory channel to reduce costs: smoke odor was triggered during the chandelier lighting sequence. The design can be simply adapted to support more channels by daisy chain networking up to 16 olfactory display devices (MIDI supports up to 128 notes that can each be used to independently drive additional olfactory channels per device). To reduce costs further, we based the design around technologies favoured by makers, such as Arduino processor boards, to build a complete system for under 300 pounds.
\end{enumerate}

For the scent we used a high-purity essential oil cade \textit{Juniperus oxycedrus} – that is known to be safely tolerated, containing naturally derived delta-Cadinene, Torreyol, Epicubenol, Zonarene and B-Caryophyllene and selected by our focus group as a potent smoke odorant evocative of burning candles).

\begin{figure} 
  \centering
    \includegraphics[width=0.8\textwidth]{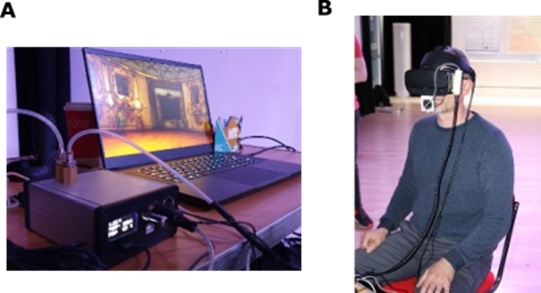} 
  \caption{University of Leicester Olfactory Display (A) alongside virtual Vauxhall Gardens laptop implementation synchronized via MIDI protocol and (B) participant experiencing the exhibition with an early olfactory display prototype (B).}
  \Description{Two photos. (A) The olfactory display device, a small box with a display next to a laptop running the VR simulation. (B) A participant wearing a VR headset and an early prototype of the olfactory display mask.}
  \label{fig:ulod_implementation}
\end{figure}

\section{EVALUATION METHODS}
\label{sec:evaluation_methods}
We first report the visitor experiences, including olfactory delivery during the public exhibition, and then, in more detail, in smaller groups, two subsequent surveys on two versions of the immersive experience, one where no odor was delivered and another one adapted to olfaction delivery.

\subsection{Exhibition's Visitor Experience}
\label{ssec:exhibition_visitor_experience}
The aim was to better understand the physiological and emotional response generated by the experience, especially the olfactory stimulus. A safe odorant was used to recreate the smell of burned candles. One frequent observation made by participants, even before experiencing the exhibit, was the uniqueness of having an olfactory component added to VR. The most common terms used to describe their experience have been engaging, immersive, and interesting. Also, when asked specifically about the olfactory experience, some of the descriptors used were candle smell, earthy smell, woody smell, or smoked paprika smell. Some of the participants also seemed to have experienced multiple odors; one of them smelled six different ones. This is despite only one odor being delivered for a short duration, when the chandelier was lit. The anticipation of there being an olfactory component in the experience appears to have heightened the sensitivity of the olfactory system to them. A physiological response was also observed for 56\% of the participants in the form of a gulp at the beginning of the odor delivery. In addition, we observed that 33\% of the participants sniffed in deeply when the odor was delivered.

\subsection{Small Group Experience without Smell Delivery (n=16)}
\label{ssec:small_group_no_smell}

\subsubsection{Presence and usability}
\label{sssec:presence_usability_no_smell}
Participants were 16 undergraduate students in Creative Computing at the University of Leicester (M=20 yo, SD=3.9 yo, 9 women, 7 men).

In the context of VR, immersion can be defined as the extent to which a VR experience engages all the senses, while presence refers to the subjective feeling of being in the virtual environment. The sense of presence is influenced by the degree of attention directed towards various elements of the virtual space and does not necessarily require complete sensory substitution [7]. 

To assess both presence and user experience--which encompasses emotions, attitudes, and ease of use  [13]--we employed focus groups and standardized questionnaires, starting with a version of the VR experience that excluded scent. 

We combined items from three validated scales: the Presence Questionnaire (PQ), the User Experience Questionnaire (USQ), and a usability questionnaire (NASA-TLX). The Presence Questionnaire, developed by Witmer and Singer; version 2] was used and has a Cronbach alpha of 0.81 in its original validation  [\cite{Witmer1998presence}, and 0.74 in our sample, indicating acceptable internal consistency. The questionnaire consists of 32 items rated on a seven-point scale. For example, item 18 asks “\textit{How compelling was your sense of moving around inside a virtual environment?}” with responses ranging from “Not Compelling” to “Very Compelling". A summary of the hypothetical factors addressed by the scale and subscales, derived from a preliminary factor analysis, is presented in Table~\ref{tab:scales_factors}. We excluded the “Haptic” questions and the corresponding haptic factor from the subscale, as the experience did not involve touch. In figures ratings are reverse-coded ratings so that higher values indicate better presence, such as reduced distraction from control mechanisms (item 24, part of Distractor Factors).

\begin{figure*} 
    \includegraphics[width=\textwidth]{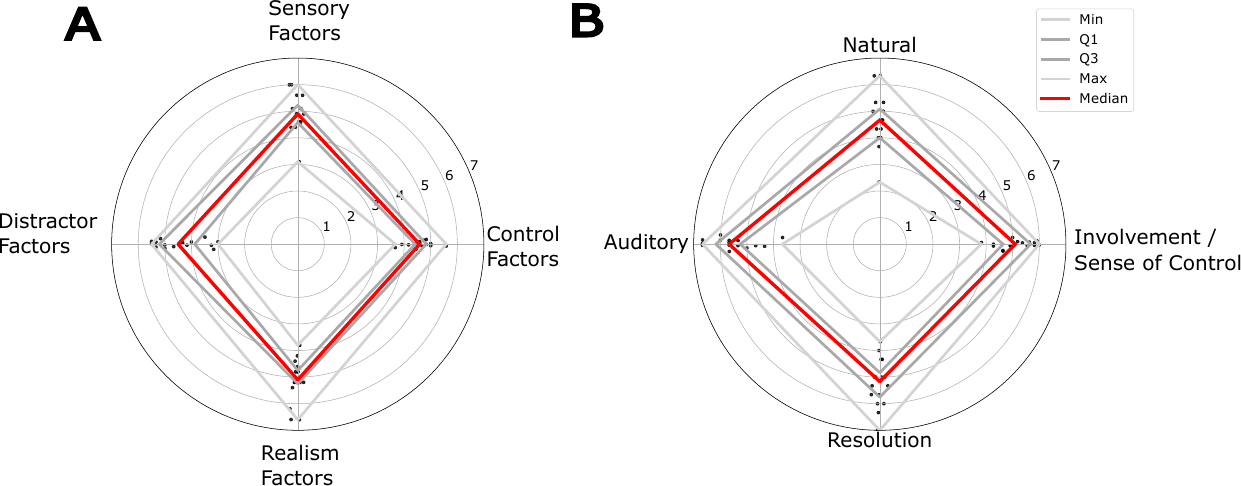} 
  \caption{Presence ratings along the dimensions of the PRESQ scale in Panel A, subscale in Panel B after participating in the VR simulation of Vauxhall Gardens. Black dots represent individual ratings. There is a small jitter applied to the coordinates to reveal individual ratings density.}
  \Description{Two radar charts. (A) Shows PRESQ ratings for Sensory Factors, Realism Factors, Distractor Factors, and Control Factors. (B) Shows PRESQ subscale ratings for Natural, Auditory, Resolution, and Involvement/Sense of Control.}
  \label{fig:presq_ueq_ratings_no_smell} 
\end{figure*}

The Presence Questionnaire (PRESQ) measured involvement in the virtual environment (e.g. “\textit{How much were you able to control events?}”), sensory experience (e.g. “\textit{How compelling was your sense of moving around inside the virtual environment?}”), distraction (e.g. “\textit{How well could you concentrate on the assigned tasks or required activities rather than on the mechanisms used to perform those tasks or activities?}") and realism ("\textit{How inconsistent or disconnected was the information coming from your various senses?}"). Participants rated their sensory presence the highest, followed by their sense of control/involvement and realism, while moderate levels of distraction were reported. Those ratings suggest opportunities for enhancing immersion across dimensions. In a subscale analysis, the auditory domain was particularly effective in generating a sense of sensory presence, as shown in Figure~\ref{fig:presq_ueq_ratings_no_smell}A.

The User Experience Questionnaire (UEQ) comprises six scales and has a Cronbach alpha ranging from 0.69 to 0.86 overall. In our sample (with one missing questionnaire, resulting in $n=15$), most scale responses showed good internal consistency (Attractiveness: 0.92, Perspicuity: 0.81, Stimulation: 0.89, Novelty: 0.86), except Efficiency (0.43) and Dependability (0.02). The discrepancy appeared to stem from item 9 (how “fast”) not correlating or correlating negatively with other items in the Efficacy scale (which included “efficient”, “practical” and “organized”), and item 8 (how “predictable”), correlating negatively with other items in the Dependability scale (which included “supportive”, “secure” and “meeting expectations"). These findings may highlight differences in dimensions that are relevant to experiencing an exhibit in contrast to experiencing a commercial product where speed and efficiency may be more positively valued.  

\begin{figure} 
    \centering
    \includegraphics[width=0.55\textwidth]{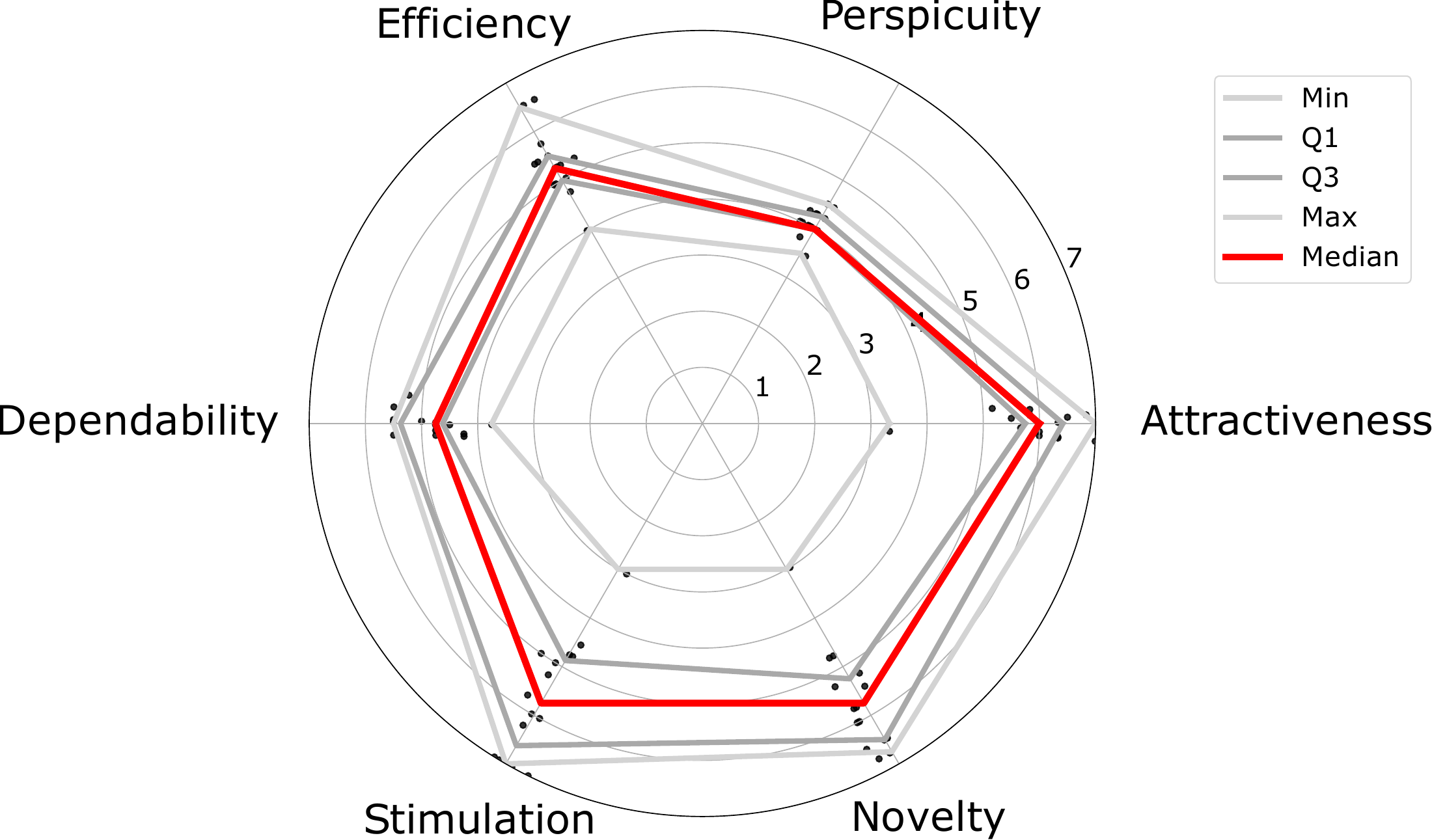} 
  \caption{Participant ratings along the dimensions of the User Experience Questionnaire (UEQ). We apply the same conventions as in Figure~\ref{fig:presq_ueq_ratings_no_smell}.}
  \Description{A radar chart showing UEQ ratings for Dependability, Efficiency, Perspicuity, Attractiveness, Novelty, and Stimulation.}
  \label{fig:ueq_ratings}
\end{figure}

Practical aspects (e.g., how intelligible is the interaction with the interface) and hedonic aspects (e.g., how stimulating is the experience) were also measured, resulting in an overall 'attractiveness' score that was exceptionally high compared to typical UEQ benchmarks, clearly leading other dimensions (Figure~\ref{fig:ueq_ratings}). Perspicuity and Novelty were also rated positively, while Efficiency and Dependability showed limitations in this experiential context. 

\begin{figure} 
  \centering
  \includegraphics[width=0.47\textwidth]{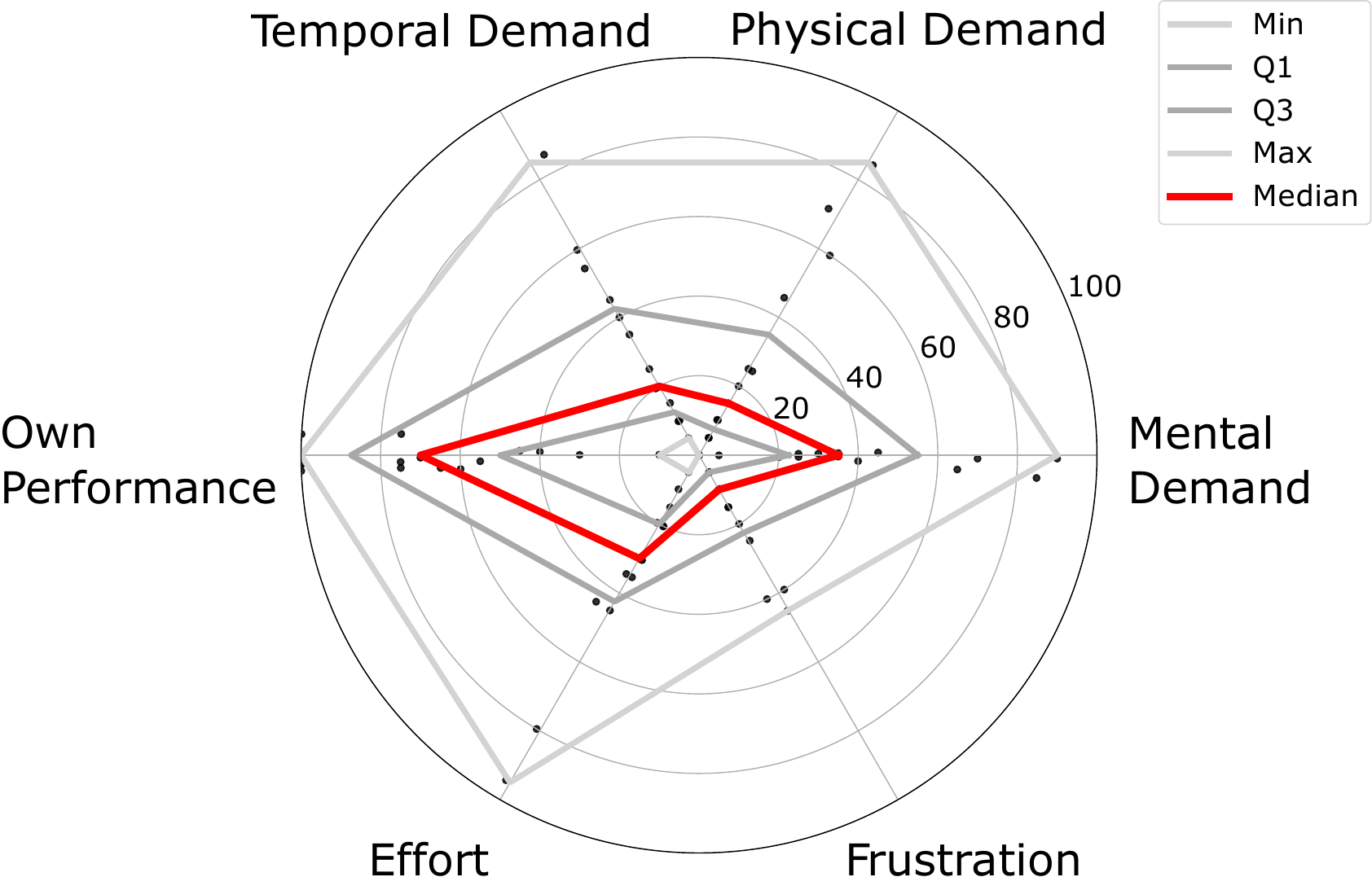} 
  \caption{Perceived level of task and workload demand on the NASA-TLX when interacting with the VR simulation of the Vauxhall Gardens. Same conventions as in Figure~\ref{fig:ulod_implementation}.} 
  \Description{A radar chart showing NASA-TLX scores for Temporal Demand, Physical Demand, Mental Demand, Effort, Frustration, and Own Performance for the no-smell condition.}
  \label{fig:nasa_tlx_no_smell}
\end{figure}

Usability refers to a multidimensional construct that gauges users' subjective level of effort and success when interacting with a device. It is particularly relevant in the context of human-computer interaction, complementing the evaluation of emotional and perceptual aspects of the VR experience. For instance, while an activity may be enjoyable and evoke a strong sense of presence, the effectiveness of the model for exploring the Rotunda can still be questioned. Given that performance goals are not well-specified in this context, we opted for the NASA-TLX [14], a well-established tool that has been used in a variety of human-computer interaction scenarios. The NASA-TLX assesses mental, physical, and temporal demands, as well as performance, effort and frustration, providing a valid measure of even in its abbreviated form. Raw scores showed very little frustration, a high sense of performance and moderate to low levels of mental, physical and temporal (e.g. how hurried the experience felt) demands (as shown by Figure~\ref{fig:nasa_tlx_no_smell}). 

\begin{longtable}{p{0.5\textwidth} p{0.4\textwidth}}
\caption{Presence, User Experience [13] and Usability Scale factors and example items.} \label{tab:scales_factors} \\
\toprule
\textbf{Presence Questionnaire (Version 2.0, PRESQ) \mbox{example items} [12]} & \textbf{Factor description} \\
\midrule
\endfirsthead
\caption[]{-- continued from previous page} \\
\toprule
\textbf{Presence Questionnaire (Version 2.0, PRESQ) \mbox{example items} [12]} & \textbf{Factor description} \\
\midrule
\endhead
\midrule \multicolumn{2}{r}{{Continued on next page}} \\
\endfoot
\bottomrule
\endlastfoot
\textit{Major factor category} & \\ 
Control factors: & \\
• \textit{How much were you able to control events?} & Degree to which the user is engaged and in control of events. \\ 
Distraction factors: & \\
• \textit{How aware were you of events occurring in the real world around you?} & Degree to which the user can concentrate on activities within the VE \\ 
\rule{0pt}{1em}
Sensory factors: & \\
• \textit{How completely were all of your senses engaged?} & Degree to which sensory inputs create a convincing VE \\ 
Realism factors: & \\
• \textit{To what degree did you feel confused or disoriented at the beginning of breaks or at the end of the experimental session?} & Degree to which the VE is seen as life-like \\ 
\midrule

\textit{Subscale} & \\ 
Involvement/control: & \\
• \textit{Were you able to anticipate what would happen next in response to the actions that you performed?} & \\  
Natural: & \\
• \textit{How much did your experiences in the virtual environment seem consistent with your real-world experiences?} & \\ 
Auditory: & \\
• \textit{How well could you identify sounds?} & \\ 
• \textit{How well could you localize sounds?} & \\ 
Resolution: & \\
• \textit{How well could you examine objects from multiple viewpoints?} & \\ 
Interface quality: & \\
• \textit{How much did the visual display quality interfere or distract you from performing assigned tasks or required activities?} & \\ 
\midrule
\textbf{User Experience Questionnaire (UEX) \mbox{example items} [12], [15]} & \textbf{Factor description} \\
\midrule
Attractiveness: & \\ 
• Annoying / enjoyable & The extent to which the product is  \\ 
• bad / good & enjoyable, attractive \\ 
• unlikable / pleasing &  \\ 
• unattractive / attractive &  \\ 
• unfriendly / friendly &  \\ 
Perspicuity: & \\ 
• not understandable / understandable & Ease with which it is easy to get familiar \\ 
• difficult to learn / easy to learn & with the product / learn  \\ 
• complicated / easy &  \\ 
• confusing / clear &  \\ 
Efficiency: & \\ 
• slow / fast & Extent to which it can be used quickly \\ 
• inefficient / efficient & and without effort   \\ 
• impractical / practical &  \\ 
• cluttered / organized &  \\ 
Dependability: & \\ 
• unpredictable / predictable & Extent to which the user feels in control \\ 
• obstructive / supportive &  \\ 
• not secure / secure &  \\ 
• does not meet expectations / meets expectations &  \\ 
Stimulation: & \\ 
• inferior / valuable & Extent to which it is enjoyable to use \\ 
• boring / exciting &  \\ 
• not interesting / interesting &  \\ 
• demotivating / motivating &  \\ 
Novelty: & \\ 
• dull / creative & Extent to which it is innovative / \\ 
• conventional / inventive & captures attention  \\ 
• usual / leading edge &  \\ 
• conservative / innovative &  \\ 
\midrule
\textbf{NASA-Task Load Index (0-100)} & \\
\midrule
Mental demand: & \\
\multicolumn{2}{p{0.9\textwidth}}{\raggedright\textit{How much mental and perceptual activity was required (e.g. thinking, deciding, calculating, remembering, looking, searching, etc.)? Was the task easy or demanding, simple of complex, exacting or forgiving?}} \\
Physical demand: & \\
\multicolumn{2}{p{0.9\textwidth}}{\raggedright\textit{How much physical activity was required (e.g. pushing, pulling, turning, controlling, activating, etc.)? Was the task easy or demanding, slow or brisk, slack or strenuous, restful or laborious?}} \\
Temporal demand: & \\
\multicolumn{2}{p{0.9\textwidth}}{\raggedright\textit{How much time pressure did you feel due to the rate or pace at which the tasks or task elements occurred? Was the pace slow or leisurely or rapid and frantic?}} \\
Own performance: & \\
\multicolumn{2}{p{0.9\textwidth}}{\raggedright\textit{How successful do you think you were in accomplishing the goal of the task set by the experimenter (or yourself)? How satisfied were you with your performance in accomplishing these goals?}} \\
Effort: & \\
\multicolumn{2}{p{0.9\textwidth}}{\raggedright\textit{How hard did you have to work (mentally and physically) to accomplish your level of performance?}} \\
Frustration: & \\
\multicolumn{2}{p{0.9\textwidth}}{\raggedright\textit{How insecure, discouraged, irritated, stressed and annoyed versus secure, gratified, content, relaxed and complacent did you feel during the task?}} \\
\end{longtable}

\subsubsection{Focus groups (2 x 5-6)}
\label{sssec:focus_groups_no_smell}
In addition to standardized scales, we gathered open reactions to the experience from a subset of the students in two focus groups (4 and 5 participants). The interviews took place in the Biomechanics and Immersive Technology Laboratory. All participants were undergraduate students in the University of Leicester's Creative Computing course and took part in the demonstration and subsequent survey and focus groups as part of their course.

To avoid leading questions the interviewer started by asking about what their feelings were about the experience and, if necessary, asked probed questions regarding the hedonic quality of the experience, ease of interaction, immersion and audio and visual experience to supplement the standardized survey.

After transcribing the discussion from audio files (\url{https://www.transcribeme.com/}), we fed transcripts into NVIVO (Version 1.7.2.). One of the authors performed a thematic analysis on the two focus groups transcripts. Codes and themes were validated by a second author. The following major themes were present in both group transcripts. Themes with the most coverage were Realism, Presence, Interaction and Enjoyment.

\textit{Realism:} Discussion of different elements that felt realistic, such as the ability to localize sounds, how the Chandelier scene in the dark felt more real as it deemphasized some of the elements that appeared at times lacking the resolution the participants were used to. The two-dimensionality of the characters, which was a scenographical choice rather than a technical one, reduced the sense of realism for some as well as inconsistent lighting when bright.

\textit{Presence:} Presence was discussed at length with active looking, the act of orienting gaze (and head) to different parts of the surrounding scene as a factor in generating a sense of being there, and for some, being transported in time. The surrounding sound was also noted as a factor improving presence. One reported a sense of warmth in the Chandelier scene, suggesting an immersive experience.

\begin{quote}
   "\textit{On the chandelier, having candles, because it just made you feel all warm because it just seemed more real.}" (P02) 
\end{quote}

\textit{Interaction:} The lack of body and scene movement (teleporting) and interaction overall within the demonstration was cited as counteracting a sense of presence in the virtual environment. The ease of use, seamlessness and clarity of the interaction were also noted.

The theme interacts with presence:

\begin{quote}
   "\textit{The more I looked around the more I felt like I was there}." (P02) 
\end{quote}

\textit{Enjoyment:} The consensus was that despite its limitations the experience was very enjoyable, it was perceived as novel despite prior experience with VR, likable, interesting and relaxing.

\begin{quote}
  "\textit{I loved it. So I loved it. I thought it was brilliant.}" (P03)  
\end{quote}

In summary, the thematic analysis supported the conclusions of the surveys and suggested that the demonstration may strike a good balance between presence, enjoyment and ease of use, as would be suitable in an exhibition context. While realism and interaction are factors that hinder presence, participants report very positive experiences and a sense of presence. It also suggests that character interaction could be an important factor. The integration of virtual agents that cross the “uncanny valley”  and Turing tests is fast becoming attainable with the progress of large-language models, crowd dynamics and biological motion synthesis.

\subsection{Small Group Experience with Smell Delivery (n=19)}
\label{ssec:small_group_with_smell}
To explore the impact of adding olfaction on the multisensory experience, we ran a second focus group as a double-blind study, with and without odor stimulus. Two sets of opaque odor column  were prepared (schematized in Figure~\ref{fig:ulod_schematic}) with and without cade oil inserts, which were then shuffled randomly. For each experimental trial one randomly selected odor column was used. Only at the end of the trial did the experimental team open the column and note down odor or no-odor stimulus condition for that trial. All other aspects of the trial we identical across the two groups.

So that the experience could be more accurately compared across individuals and conditions for this study, we adapted the VR implementation to remove exploration in space (linear narrative), although freedom of head movement was retained. This also fixed its duration, which simplified the synchronization of the odor delivery and allowed us to ask users to estimate the duration of the VR experience.

Key information subjects were told at the start of each trial:

\begin{quote}
    \textit{The aim of this study is to evaluate the impact of multi-sensory stimuli on measures of users' perceptions and attitudes towards the exhibit. You will be asked to wear a VR headset to visit the exhibition. During your visit, your progression will be guided/narrated as you experience the sights, sounds and/or smells that are designed to recreate different aspects of Pleasure Gardens.}
\end{quote}

Because the odor delivery set-up with the VR headset was obvious to subjects at the start of the trial, this may have created an expectation of odor delivery for both groups, especially as the odor display was activated synchronously with the lighting of the candles at a single point in the linear narrative (duration $\sim$15 seconds).

\subsubsection{Presence, immersive tendency and usability}
\label{sssec:presence_usability_with_smell}

There were 19 participants of ages ranging from 21 to 65 years old (M = 37 yo, SD=15 yo) of diverse backgrounds from our university population consenting to the study (see Ethics). For practical reasons, as it took longer to set up the demonstration, including smell delivery, we had a short version of the PRESQ (item 4, 6, 9, 10, 11, 12, 13, 19, 20). We selected items that loaded strongly on hypothetical presence factors and were relevant to the experience. We combined this with a shorter version of the Immersive Tendency Questionnaire (ITQ, Version 2.0), with items 3, 4, 5, 6, 7, 16 and 17, to be able to examine the role of inter-individual differences, in particular the ability to deeply focus and be deeply involved in an experience, such as while watching TV or reading a book.

The shorter version of the PRESQ had a Cronbach's alpha of 0.78 in our sample. The shorter version of the ITQ had a Cronbach's alpha of 0.95 in our sample, suggesting that the items we picked contributed strongly to the measured construct. We evaluated the task load again using the NASA-TLX questionnaire as above. We were specifically interested in seeing whether smell delivery and/or the affected immersion related to what we have seen above. Correlation between the ITQ z-score and PRESQ z-score in the sample was moderate to high, Pearson's $r = 0.59$, $p<0.01$, suggesting that presence is enhanced in people who are more easily drawn into various types of media.

\begin{figure*} 
  \includegraphics[width=0.9\textwidth]{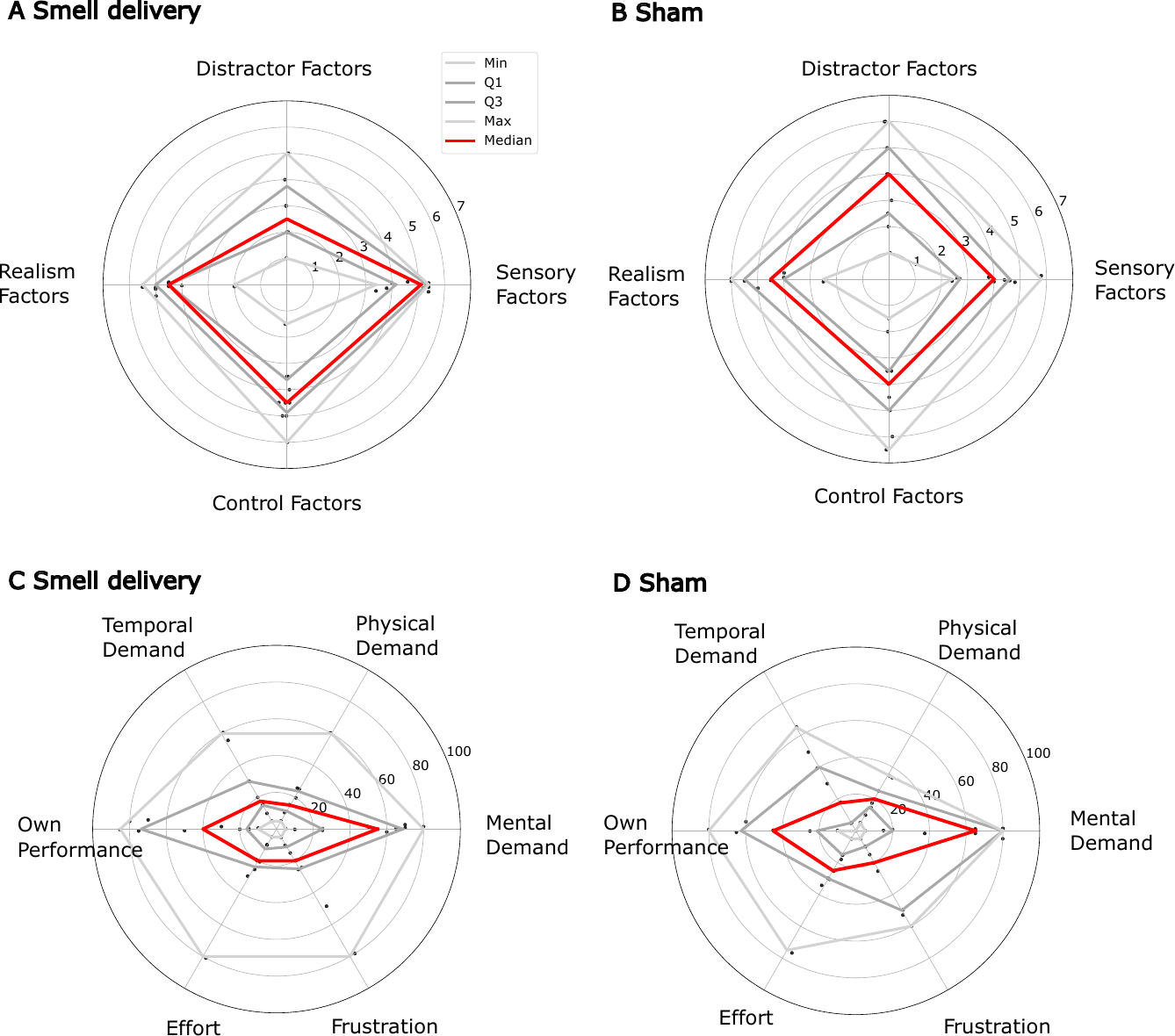}
  \caption{Panel A shows PRESQ ratings when smell was not delivered (n = 6). Panel B shows PRESQ ratings when the smell-delivery system was set up but no smell was delivered (n=14). Panel C-D shows the corresponding NASA-TLX ratings. Plotting follows the same conventions as in Figure 7.} 
  \Description{Four radar charts. (A) PRESQ ratings for smell delivery condition. (B) PRESQ ratings for the sham (no smell) condition. (C) NASA-TLX ratings for smell delivery condition. (D) NASA-TLX ratings for sham condition. All charts show dimensions like Distractor Factors, Realism Factors, Sensory Factors, Control Factors for PRESQ, and Temporal Demand, Physical Demand, Mental Demand, Own Performance, Effort, Frustration for NASA-TLX.}
  \label{fig:presq_nasa_tlx_with_smell}
\end{figure*}

Figure~\ref{fig:presq_nasa_tlx_with_smell} shows a similar pattern of response across both groups with the distractor factor lower and lower spread to scoring of the sensory factors in the odor delivery group. Demand was found to be quite equivalent for the two groups. As in the previous experiment, higher values indicate better immersion, including for Distraction Factors after recoding. Overall, we can see that participants reported high levels of presence and similar patterns across factors as in the first demonstration without the smell delivery system. On average, people were more aware of the display device (the Distractor Factors item selected) in the smell delivery group, but differences between them were not statistically significant (independent samples t-tests ps > 0.193; uncorrected for multiple comparisons). Further studies would be needed to confirm this pattern.

Due to the fixed length of the experience, we asked participants to judge the time duration. Overall, both groups overestimated the actual duration (ground truth value = 5.7 min) with a shorter estimate for the odor group, although estimates had wide variation across individuals (odor 6.5 $\pm$ 2.5 min and no-odor 7.3 $\pm$ 2.5 min). For each trial, we played an audio clip of a telephone ringing in the laboratory at the same time point in the narrative and found no significant difference in the noticing of this ($\sim$60\% in both groups positively reported).

A significant result is that in the odor group, there was large variation in reporting any odor stimulus present at all which might be explained by individual differences in sensory thresholds and specific anosmias. Intriguingly, this expectation apparently led to some false positive odor reports in the no-odor group, with respondents noting:

\begin{quote}
\textit{“There was an artificial approximation of fresh air during the first part where there were pillars appearing above the blue floor”} (N02 – no odor group)
\end{quote}

\begin{quote}
\textit{“Faint wood smoke. Not sure when the wood smoke started - became gradually aware of it - just before the chandelier” }(N06 – no odor group)
\end{quote}

We also found some participants in the odor group describing olfactory experiences that we not present:

\begin{quote}
\textit{“A gust of cold/fresh air as the simulation transitioned from the opening water scene to the grass field”} (N10 – odor group)
\end{quote}
\subsubsection{Content analysis}
\label{sssec:content_analysis_with_smell}
We asked participants to describe the experience in 3 words. We analyzed the frequency of words being used and organized them in categories in Table~\ref{tab:three_word_descriptors}. There is a strong correspondence between the statements and what came out of the focus groups without smell delivery. This is important, as smell thresholds are very different across people. Note that all participants thought that smells were being delivered but they were only delivered to some. Sometimes participants reported smells that were not delivered. We were not necessarily expecting a positive experience in this pilot experiment:what is pleasant to some can be considered very intrusive and distracting to others. The delivery also relies on the addition of a device that could have hindered the experience.

\begin{table} 
\centering
\caption{Counts of in response to “Use three words to describe the VR exhibit “, organized in categories.}
\label{tab:three_word_descriptors}
\begin{tabular}{lp{5cm}c}
\toprule
CATEGORY & Words & Counts \\
\midrule
AESTHETICS & beautiful (2), elegant (1), surreal (1) & 4 \\
EMOTION & calm (2), relaxing (3), enjoyable (1), pleasant (1), fun (2), entertaining (1) & 10 \\
INSIGHT & informative (6), factual (1), historical (1), interesting (6), imagination-provoking (1), narrative (1) & 16 \\
PRESENCE / IMMERSION & immersive (4), engaging (2), large (the architecture) (1) & 7 \\
NOVELTY & innovative (1), novel (2), new (1), obscure (1) & 5 \\
SIMPLICITY & easy (1), clear (1), brief (1), short (1), static (1), non-interactive (1), detailed (1) & 7 \\
\bottomrule
\end{tabular}
\end{table}

\section{DISCUSSION}
\label{sec:discussion}

The Virtual Vauxhall Gardens (VVG) project highlights key aspects of designing and evaluating multisensory VR heritage experiences. Our baseline visual-auditory VR reconstruction effectively immersed users, with evaluations indicating strong auditory presence and user control, alongside an exceptionally high degree of perceived attractiveness (UEQ). Qualitative feedback affirmed this positive experiential quality and a desire for greater interactivity, though it also noted technical limitations like display resolution. Crucially, these initial evaluations underscored the challenges of applying standard metrics: usability scales like UEQ’s Efficiency and Dependability showed poor reliability in this non-task-oriented context, and PRESQ's ‘realism’ factor proved less relevant than evocative engagement for our interpretative historical reconstruction.

The integration of synchronized olfaction demonstrably amplified user engagement. This was evidenced by qualitative descriptors such as "engaging" and "immersive," and by physiological reactions like gulping or sniffing at scent delivery. The novelty of the olfactory element was a notable factor. Intriguingly, the mere anticipation of smell appeared to heighten overall sensory awareness, occasionally leading to participants reporting scents that were not actually delivered. While rigorous quantification of olfaction's impact on standardized presence scores necessitates larger-scale studies, our qualitative data indicate that the olfactory dimension contributed to a uniquely memorable and positive encounter with the historical setting.

\subsection{Olfaction and Presence}
\label{ssec:vvg_literature_olfaction_presence}
Our findings on olfaction's impact resonate with a complex body of research on its link to presence in virtual environments. While some studies suggest congruent scents can increase reported presence [1, 16, 17], this effect is not universal and is highly context-dependent [1]. For instance, interactivity or the emotional tone of a scene can be stronger drivers than olfaction alone, and specific scents may not always enhance immersion [1, 18]. Moreover, presence itself may not be sufficient for achieving certain outcomes like therapeutic efficacy [19]. Congruence between olfactory, visual, and auditory stimuli, as employed in VVG, appears crucial and can improve recall [20], although the nature of the scent (pleasant vs. unpleasant) can also modulate its effect on presence.

However, integrating additional sensory modalities risks increasing cognitive load or sensory overload, potentially hindering other aspects of the experience, such as recall in learning tasks or task awareness [21, 22]. While VVG participants responded positively to the experiential tour, suggesting enhanced engagement, this might not translate to applications requiring sustained focus. The positive emotional engagement, aligning with findings on calm and relaxation [18, 23], and the influence of anticipation and novelty on the VVG olfactory experience, are noteworthy. This suggests that in novel exhibition settings, user priming can heighten sensory awareness, making the olfactory dimension subjectively impactful [cf. 24 for aromatherapy/VR].

\subsection{Persistent Technical Challenges}
\label{ssec:vvg_successes_context}
The development of the ULOD represents a pragmatic technical solution for temporary museum exhibitions, demonstrating feasible synchronized olfaction. The novel use of MIDI for synchronization addressed challenges in timing scent delivery with VR events [25], and its low cost offers an attractive option for institutions with limited budgets [26]. Positive user feedback suggests this solution created a compelling multisensory experience within VVG's scope.
Despite VVG's functional system, persistent technical challenges hinder widespread, sophisticated olfactory displays in VR [27]. The ULOD sought to mitigate common issues like contamination, safety, cost, and synchronization, yet these remain broader hurdles. Key challenges include the integration of bulky with HMDs [28, 27]; complex scent generation and precise diffusion control [27]; achieving rapid and precise temporal synchronization for dynamic events [29]; cross-contamination, especially with lipophilic odorants (VVG used a single channel and PTFE to mitigate this) [29]; cost and accessibility [27]; ensuring chemical safety and accommodating individual olfactory variability--such as for users who may have allergies or chemical sensitivities; and maintenance logistics for public deployment [26, 17].

The VVG project's technical implementation is a valuable case study for affordable, triggered olfactory cues in specific heritage applications, with MIDI offering a precise, low-latency, and networkable solution. However, realizing rich historical smellscapes beyond single-triggered events awaits further breakthroughs, especially in multi-scent delivery fidelity, contamination control, and seamless hardware integration for highly interactive scenarios [29].

\subsection{Challenge of Recreating Historical Sensory Experiences }
\label{ssec:vvg_reconstruction_interpretation}
There were several challenges in the recreation of VVG. While visual/auditory aspects drew on archival sources like Wale's 1752 print, interpretive leaps were needed for missing details. This was particularly true for olfaction, where, acknowledging differing modern sensibilities, we focused on an evocative, event-linked scent (chandelier lighting) rather than attempting a literal recreation of potentially unpleasant ambient historical smells.

\subsection{The Challenge of Recreating Historical Sensory Experiences}
\label{ssec:challenge_historical_sensory} 
Reconstructing the sensory world of the past presents profound challenges, particularly when dealing with the sense of smell. Odors are inherently ephemeral, leaving little direct physical evidence behind. Our modern olfactory language is often imprecise and reliant on culturally specific metaphors, making it difficult to accurately describe or compare historical smells. Furthermore, the meaning and perception of odors are highly contextual, shaped by specific cultures, times, and places, and contemporary olfactory sensibilities may differ significantly from those of the past. This combination of factors—ephemerality, linguistic limitations, context-dependency, and perceptual shifts—makes accessing and recreating the olfactory dimension of history exceptionally difficult [30].
The VVG project aligns with developing fields of sensory history and digital heritage, which emphasize structured methodologies for documenting ephemeral sensory data like odors as cultural heritage [30] and utilize technologies like VR for their reconstruction.

The concept of ‘authenticity' in digital heritage is multifaceted. As VVG aimed to capture "something of what it might have felt to be there," our approach aligns with scholarship suggesting authenticity extends beyond photorealism to include narrative coherence, contextualization, and emotional response [31]. Multisensory engagement can enhance this, even with interpreted stimuli [32], and aids in preserving intangible cultural heritage like atmosphere or sensory experiences [33], though ethical considerations of interpretation and accessibility remain.

The VVG project exemplifies the interplay between research and creative interpretation necessary for sensory heritage. Our use of an evocative, event-linked scent reflects a pragmatic adaptation to contemporary requirements and the impossibility of perfectly replicating past smells. This supports the view that authenticity here stems from plausible, research-informed representations involving interpretation, rather than flawless imitation [31]. Transparency in detailing sources, choices (e.g., "educated conjecture" for colors), and rationale, as practiced in VVG, is crucial to allow users to distinguish fact from informed interpretation [30].

\subsection{ Evaluation of Visitor’s Experience }
\label{ssec:vvg_evaluation_findings_challenges}
In addition to qualitative evaluation, the use of standardized scales allowed a more systematic exploration of the visitor experience. Yet, the limitations of standardized questionnaires in this context were evident: PRESQ's 'Realism' factor was ill-suited for VVG’s interpretative elements, and certain UEQ scales (Efficiency, Dependability) showed poor reliability, highlighting challenges in applying generic user experience instruments to unique heritage experiences. This suggests a need for bespoke evaluation approaches in VR digital heritage. Qualitative data from focus groups, however, proved invaluable, complementing quantitative scores by revealing themes of Realism, Presence (including unexpected sensations like 'warmth'), Interaction desires, and overall Enjoyment, offering nuanced insights into the user experience. VVG's mixed-methods evaluation aligns with current trends addressing the limitations of questionnaires for subjective presence [34, 19]. Increasingly, studies triangulate self-reports with objective physiological or behavioral measures (e.g., biosensors, EEG, observational data) for a more holistic assessment of user experience [35, 36]. Evaluating multisensory integration, particularly olfaction, adds complexity, with studies using varied approaches to probe its impact on presence, awareness, and other constructs [21, 25, 8]. In VVG qualitative and questionnaire data was complemented with novel (albeit preliminary) physiological observations, from sniffing behaviour. Importantly, we contend that heritage evaluation must seek to optimize different goals than mere presence or efficacy; for instance focusing on engagement, learning, and authenticity, not task efficiency or resolution, which could be appropriate in other such as gaming [31].

\subsection{Limitations and Future Directions}
\label{ssec:future_directions}
Immersive virtual reality (VR) systems have the potential to enhance our understanding of past sensory experiences [39].  By integrating visual, auditory and smell, moving beyond a traditional focus on sight alone, we can further enhance the emotional connection with historical experiences. Our research emphasizes the need for advancements in olfactory technology, the development of evaluation methodologies tailored for multisensory experiences, and the importance of interdisciplinary collaboration among researchers from various fields. Long-term impact studies could assess the effects of these experiences on learning and emotional connections with cultural heritage. By using technologies like GIS, 3D mapping, and VR/AR, and multiple senses to reconstruct past experiences, we can acknowledge the complexity of human perception and interaction with historical environments.

Technically, continued innovation in odorant delivery is needed to create more robust, reliable, scalable, affordable, and user-friendly olfactory display systems capable of delivering a wider range of scents with high fidelity, minimal contamination, and precise temporal control.
One intriguing aspect of the exhibition, which would bear further investigation, is the extent to which the VR environment offers a pleasurable and even therapeutic experience for those with certain conditions. Informal discussions revealed that a user with a serious balance disorder experienced an improvement in proprioceptive (awareness of body position) and vestibular (balance) function while immersed in the environment and subsequently. Furthermore, autistic users reported that the sensory elements (including smell) were pleasurably stimulating, rather than the expected overload. These comments were anecdotal but sufficiently striking to warrant further study. No doubt pacing and user control were essential components of the experience, but also the anticipation of pleasure created by the historical framing of the installation played a part. Technological advances in VR may be able to help in shaping the sensory experience to the individual user by adjusting to deficiencies in optics or other areas [40]. 
Some users commented on the lack of physical movement (the experience was entirely sedentary), and any future attempt to realize the whole Gardens would need to include some aspects of movement. It would also enhance the model to enable active conversation with the characters represented. This was part of the original plan, but could not be realized due to time and budgetary constraints.

\subsection{Conclusion}
The VVG project explored the potentially transformative power of multisensory virtual reality in cultural heritage. By weaving together sight, sound, and the evocative sense of smell, we sought to enhance our understanding of historical experiences and deepen emotional connections to the past. Visitors rich tapestry of experiences confirmed the appeal of the approach. We highlight challenges ahead--technical, methodological, and ethical--which require the type of massively multidisciplinary approach that was implemented in this project.

\section{ETHICS AND DATA AVAILABILITY}
User surveys were approved by the School of Psychology Ethics Committee. The anonymized raw data is publicly available on the Open Science Framework website (\textit{URL to be added}). Data and software are available upon request. 

\begin{acks}
Funding was secured from the University's Knowledge Exchange and Enterprise Development Fund (KEEF) and from the Leicester Institute of Advanced Studies to support the creation of an initial model. For the purpose of open access, the author has applied a Creative Commons Attribution license (CC BY) to any Author Accepted Manuscript version arising from this submission.
\end{acks}


\end{document}